
\raggedbottom
\tolerance=100000
\hyphenpenalty=100000
\magnification=\magstep1

\def\frac#1/#2{\leavevmode\kern.1em
\raise.5ex\hbox{\the\scriptfont0 #1}\kern-.1em
/\kern-.15em\lower.25ex\hbox{\the\scriptfont0 #2}}
\def\lsim{\, \lower2truept\hbox{${<
\atop\hbox{\raise4truept\hbox{$\sim$}}}$}\,}
\def\gsim{\, \lower2truept\hbox{${>
\atop\hbox{\raise4truept\hbox{$\sim$}}}$}\,}

\def\ninit{\hoffset=.0 truecm
            \voffset=.0 truecm
            \hsize=15.5 truecm
            \vsize=23.5 truecm
            \baselineskip=20.pt
            \lineskip=0pt
            \lineskiplimit=0pt}

\def\pag{\pageno=2\footline={\hss\tenrm\folio\hss}}
\def\pag{\pageno=2\headline={\hss\tenrm\folio\hss}}
\ninit
\def\oneskip{\vskip\baselineskip}       
\def\ref{\noindent\hangindent=20pt\hangafter=1}
\centerline{\null}
\nopagenumbers
\noindent
\centerline
{\bf ANGULAR CORRELATIONS OF THE X--RAY BACKGROUND}
\centerline {\bf AND CLUSTERING OF EXTRAGALACTIC X--RAY SOURCES}
\oneskip
\parindent=25pt
\parskip 0pt
\centerline{{\it L. Danese$\,{}^1$}, {\it L. Toffolatti$\,{}^2$},
{\it A. Franceschini$\,{}^2$},}

\centerline{{\it J.M.
Mart\'{\i}n-Mirones$\,{}^{3,2}$} and {\it G. De Zotti$\,{}^2$}}

\medskip\noindent
\llap{${}^1$}Dipartimento di Astronomia, Vicolo dell'Osservatorio 5,
I--35122 Padova, Italy

\medskip\noindent
\llap{${}^2$}Osservatorio Astronomico, Vicolo dell'Osservatorio 5,
I--35122 Padova, Italy

\medskip\noindent
\llap{${}^3$}Departamento de F{\'\i}sica Moderna, Universidad de Cantabria,
Avda. de Los Castros s/n, E--39005 Santander, Cantabria, Spain
(permanent address)

\def\aa #1 #2/{{ A\&A}, { #1}, #2}
\def\araa #1 #2/{{ ARA\&A}, { #1}, #2}
\def\aj #1 #2/{{ AJ},{ #1}, #2}
\def\apj #1 #2/{{ ApJ},{ #1}, #2}
\def\apjl #1 #2/{{  ApJ (Letters)}, { #1}, L#2}
\def\apjs #1 #2/{{ ApJS}, { #1}, #2}
\def\mnras #1 #2/{{ MNRAS}, { #1}, #2}
\def\qjras #1 #2/{{ QJRAS}, { #1}, #2}
\def\pasp #1 #2/{{ PASP}, { #1}, #2}
\def\nat #1 #2/{{ Nature}, {\ #1}, #2}
\def\physl #1 #2/{{ Phys.Lett.}, { #1}, #2}
\def\physrep #1 #2/{{ Phys.Rep.}, { #1}, #2}
\def\physrev #1 #2/{{ Phys.Rev.}, { #1}, #2}
\def\physrevb #1 #2/{{ Phys.Rev.B}, { #1}, #2}
\def\physrevd #1 #2/{{ Phys.Rev.D}, { #1}, #2}
\def\physrevl #1 #2/{{ Phys.Rev.Lett.}, { #1}, #2}
\def\sovastr #1 #2/{{ Sov.Astr.}, { #1}, #2}
\def\sovastrl #1 #2/{{ Sov.Astr. (Lett.)}, { #1}, L#2}
\def\commastr #1 #2/{{ Comm.Astr.}, { #1}, #2}
\def\book #1 {{\it ``{#1}'',\ }}

\vfill\eject

\smallskip
\pag
\oneskip
\oneskip
\centerline {\bf ABSTRACT}
\oneskip
The information content of the autocorrelation function (ACF) of intensity
fluctuations of the X--ray background (XRB) is analyzed. The tight upper
limits set by ROSAT deep survey data on the ACF at arcmin scales imply
strong constraints on clustering properties of X--ray sources at
cosmological distances and on their contribution to the soft XRB. If quasars
have a clustering radius $r_0 = 12$--20 Mpc ($H_0 = 50$), and their
two point correlation function, $\xi(r)$, is constant in comoving
coordinates ($\epsilon =-1.2$), as indicated by optical data, they
cannot make up more 40--50\% of the soft XRB (higher contributions
corresponding to lower $r_0$); the maximum contribution may reach 80\%
in the case of stable clustering ($\epsilon = 0$).
If $r_0 \geq  12\,$Mpc, a slow decrease
of the $\xi(r)$ of the AGNs with increasing redshift ($\epsilon \simeq -3$) is
ruled out since it would imply an implausibly low contribution
to the XRB. Active Star-forming  (ASF) galaxies clustered like normal
galaxies ($r_0 \simeq 12\,$Mpc) can yield up to 20\% or up to 40\% of
the soft XRB for $\epsilon = -1.2$ or $\epsilon = 0$, respectively.

The ACF on degree scales, typical of existing hard X--ray surveys,
essentially reflects the clustering properties of local sources and is
proportional to their volume emissivity. The upper limits on scales of a few
degrees imply that hard X--ray selected AGNs have $r_0 \leq 25\,$Mpc if
$\epsilon = 0$  or $r_0 \leq 20\,$Mpc if $\epsilon = -1.2$. No significant
constraints are set on clustering of ASF galaxies, due to their low
local volume emissivity. The possible signal on scales $\geq 6^\circ$,
if real, may be due to AGNs with $r_0 \simeq 20\,$Mpc; the contribution
from clusters of galaxies with $r_0 \simeq 50\,$Mpc is a factor
$\simeq 2$ lower. Other classes of sources clustered like normal
galaxies could only account for such signal if their local volume emissivity
is $\approx 6\times 10^{38}\,\hbox{erg}\,\hbox{s}^{-1}\,\hbox{Mpc}^{-3}$.
This value is somewhat in contrast with larger estimates based on alternative
methods, and implies that the bulk of the XRB is not local.

We have also computed the expected ACF in the 2--10 keV energy band
on arcminute scales, that will be useful for comparison
to the soon coming data from the ASTRO--D satellite.

Although the nature of sources producing the bulk of the soft and of the
hard XRB is likely to be different, their clustering properties
appear to be not much different from those of normal galaxies.

\medskip\noindent
{\it Subject headings:} cosmology --- galaxies: clustering --- quasars
--- X--ray: general

\parindent=0.7truecm
\vfill\eject

\oneskip
\centerline {1. INTRODUCTION}
\oneskip\nobreak
A good deal of information on the origin of the X--ray background (XRB) as
well as on the evolution of clustering in the Universe is imprinted in
the angular distribution of the XRB. Studies of fluctuations and of their
AutoCorrelation Function (ACF) as well as of the cross--correlation
of the XRB with source populations selected in optical, IR and
radio bands are useful in constraining counts, evolution
and clustering properties of extragalactic X--ray sources.

Marshall et al. (1980) pointed out that their data on the spectrum
of the XRB in the energy interval 3-60 keV (henceforth HXRB) are well
described by thermal bremsstrahlung emission at a temperature
$T\simeq 40$ keV. On the other hand the spectrum of the XRB below 3 keV
(henceforth SXRB) is still subject of debate (see e.g. McCammon \& Sanders,
1990). Nevertheless it is generally accepted that the SXRB exceeds the
extrapolation of the HXRB (Wu et al. 1990; Micela et al. 1991; Hasinger 1992)
and that at energies lower than 1 keV the galactic contribution is
increasingly important.

Only at soft energies ($E\leq 3$ keV) X--ray imaging
capabilities have already attained resolutions of few tens of arcsecond,
while
at harder energies collimators allow resolutions of few degrees at most.
Correspondingly, at soft energies it has been possible to detect sources
fainter by almost 4 orders of magnitude than those detected at hard energies.
For instance, ROSAT Deep Surveys show that about 40$\%$ of the SXRB
is contributed by sources brighter than S(0.5--2 keV) $\geq 7\times 10^{-15}$
$erg\ s^{-1}cm^{-2}$ (Hasinger, Schmidt \& Tr\"umper 1991; Shanks
et al. 1991; Anderson et al. 1992), while only a few percent
of the HRXB has already been resolved into sources.
However it is worth noticing that in the next future high resolutions
will also be available at higher energies; in particular ASTRO--D is expected
to
produce maps with resolution of about 2$'$ in the 0.5--10 keV energy range
(Inoue 1992).
On the other hand there is increasing evidence that the spectral properties
of the
source populations detected by deep surveys at soft energies are rather
different
from those required to account for the HXRB
(see Franceschini et al. 1992, for a detailed discussion).
For these reasons in the following we will discuss the
autocorrelations of the SXRB and of the HXRB separately.

The XRB ACF is an integrated view of the
clustering properties of the source populations contributing to the XRB.
Thus studies of autocorrelations provide an important tool to explore
clustering
evolution since the formation of structures in the Universe (Wolfe \&
Burbidge 1970; Schwartz et al. 1971; Fabian 1972).
Depending on energy bands, on
limiting fluxes and on angular scales, the clustering properties of
different X--ray source populations can be elicited.
On the other hand the clustering properties of galaxies, galaxy clusters and
QSOs have been studied in optical, far-IR and radio bands. Therefore it is
quite informative to compare the expected contribution of these populations
to the observed ACF of the XRB, looking for constraints
on clustering and/or emissivity of the sources.

Studies on ACF of the XRB have been worked out by many
authors, exploiting the different available data bases.

As for the SXRB, Barcons and Fabian (1989) derived the ACF on
angular scales in the range $1^{'}$ to $15^{'}$, using extremely deep
exposures taken with the Imaging Proportional Counter (IPC) on board of the
Einstein Observatory. Their nominal energy band is 1-3 keV.
On similar scales and energy band (0.8--3.5 keV) and using
again extremely deep IPC exposures Soltan (1991) has found
quite stringent upper limits to the ACF on angular scales ranging from
2 to 4 arcminutes. More recently deep pointed
ROSAT observations showed that the SXRB in the 0.5--2 keV band
is actually very smooth on arcminute scales.
(Hasinger, Schmidt \& Tr\"umper 1991;
Hasinger 1992; Georgantopoulos et al. 1991; Carrera \& Barcons 1992).
The ensuing limits on the clustering as well as
on the contribution to the SXRB of the X--ray sources are
quite interesting and competitive with the limits obtained from
optical surveys (see Danese, De Zotti \&
Andreani, 1992 for a review).

Concerning the ACF of the HXRB,
Persic et al. (1989) have derived upper limits for
angular separations ranging from $3^{\circ }$ to $27^{\circ }$
by analyzing  HEAO~1 A-2 data in the nominal 2--10 keV band; the ensuing
constraints have been discussed by De Zotti et al. (1990).
Carrera et al. (1991) explored the ACF on similar energy interval
(4-12 keV) and angular scales using exposures
obtained with the Large Area Counter (LAC) on board of the $Ginga$ satellite
and found results consistent with those of Persic et al. (1989).
Mart\'{\i}n-Mirones et al. (1991) and Carrera et al. (1992) studied the
ACF on sub-degree angular scales using A-2 and LAC data respectively.
Again the results of both analyses are fully consistent and have
been used to put significant constraints on clustering of AGNs and galaxy
clusters.

Extending the analysis to a larger portion of the sky covered by the
A-2 experiment,
Mushotzky \& Jahoda (1992) have reported a possible detection of positive
autocorrelation at scales in the range $6^{\circ }$ to $20^{\circ }$.

In this paper we will exploit the whole body of data to examine the
limits on clustering, clustering evolution and volume emissivity of X--ray
sources contributing to the SXRB and HXRB. In Section 2 the theory
of the ACF is presented and the effects of the characteristics of individual
experiments on the results are briefly discussed. In Section 3 and 4
we present the constraints on emissivity and clustering properties
of AGNs and galaxies following from the current limits on ACF of the SXRB
and HXRB respectively. Section 5 is devoted to the discussion and conclusions.

A Hubble constant $H_{0}=50$ and a cosmological deceleration parameter
$q_0=0.5$ are used throughout the paper.

\oneskip
\oneskip
\centerline {2. THE ANGULAR AUTOCORRELATION FUNCTION}\nobreak
\oneskip\nobreak
Cell-to-cell fluctuations of randomly distributed or clustered
unresolved sources produce intensity fluctuations
$\delta I({\bf n})=I({\bf n})-\langle I\rangle $
of the observed background. The angular ACF 
of the intensity fluctuations is usually defined as

$$ W(\theta)={\langle\delta I({\bf n})\delta I({\bf n'})
\rangle \over \langle I\rangle ^2},
\eqno (1)$$

\noindent
where $\theta $ is the angle between the directions ${\bf n}$ and ${\bf n'}$,
and the spatial average $\langle I\rangle $ is referred to the {\it residual
background}, once the detected sources have been subtracted.
The contribution to the ACF from clustered source populations can
be computed under rather general hypotheses (see e.g. Mart\'{\i}n-Mirones
et al. 1991). If the maximum clustering scale
$r_{max}$ is much smaller than the Hubble radius,
$c/H_{\circ }$, and the maximum value of angular separation is
much less than one radian, then
the proper separation $r$
between points on two lines of sight separated by an angle $\theta $ can
be approximated by:

$$r = \left[ (c \delta t)^2 + (d_A \theta )^2 \right]^{1/2}, \eqno(2)$$
where $d_A = d_L (1 + z)^{-2}$ is the
angular diameter distance,
and $\delta t = (dt/dz)\delta z = - H_o^{-1}
(1 + z)^{-2} (1 + \Omega z)^{-1/2} \delta z$.
The luminosity distance is given by $ d_L = (2c/\Omega^2 H_0) \{
\Omega z + (\Omega - 2) [ -1 +
(\Omega z + 1)^{1/2} ] \}$.

With the additional hypothesis that the beams
[response function $f(\vartheta , \varphi )$] do not overlap,
Mart\'{\i}n-Mirones et al. (1991) have shown that the ACF as a function of the
angular separation $\theta $ is given by:

$$ \eqalign { \hfill W(\theta  )=  {\left(c \over 4 \pi H_o
{\langle I\rangle }\right)^2}
\int d\omega f(\vartheta , \varphi ) \int d\overline \omega
f(\overline \vartheta  , \overline \varphi ) \cdot \hfill \cr
\hfill \cdot \int_{z_m(L_{min},S_l)}^{z_{max}} dz  {j_{eff}^2(z)
\over (1 + z)^4 (1 + \Omega z)} \int_{\max \left[z_m - z, -\Delta
(r_{max})\right] }^{\min \left[z_{max} - z, \Delta(r_{max})\right] }
d(\delta z) \xi (r,z) ,\hfill \cr} \eqno(3)$$
where $\Delta (r_{max})$ is the value of $\delta z$ corresponding to the
maximum scale of clustering, and
$$j_{eff}(z) = \int_{L_{min}}^{min\left[ L_{max}, L(S_l,z) \right] } d\log L
\ L \ n_c(L,z) K(L,z) \eqno(4)$$
is the effective volume emissivity; $L(S_l,z)$ is the luminosity of a source
at the
redshift $z$ that yields a flux equal to the detection limit $S_l$,
whereas $z_m(L_{min},S_l)$ is the redshift at which a source of minimum
luminosity $L_{min}$ has the limiting flux $S_l$ and $z_{max}$ is the
upper limit to the redshift of sources. The average
$\langle I \rangle $ is performed after subtracting sources
brighter than $S_l$.

The epoch--dependent Luminosity Function of the clus\-ter\-ed sources is
$n_c(L,z)\,d\log L$, $\xi (r,z)$ is the epoch--dependent spatial
cor\-re\-la\-tion
function and the K-correction factor is given by
$$K(L,z) = {\int^{E_2}_{E_1} L[E(1+z),z]\, dE \over
\int^{E_2}_{E_1} L(E,0)\, dE}, \eqno(5)$$
$E_1$ and $E_2$ bounding the relevant energy band.

Available data suggest that $\xi (r,z)$ can be represented as (see Peebles
1980;
Bahcall \& Soneira 1983; Sebok 1986)

$$\xi (r,z)=D^2(z)\xi_0 (r)\eqno (6)$$
$$\xi _0={\left(r_0/r\right)}^{\gamma }\eqno (7)$$
with $\gamma \simeq 1.8$ for a number of populations of extragalactic sources.

Actually, in numerical calculations a more refined model is used, that
make the correlation function flat at small radii and explicitly
incorporate a maximum radius $r_{max}$, such as $\xi =0$ if $r >r_{max}$.
In the following we will adopt $r_{max}= 3 r_0$.

The factor $D^2(z)$ allows for clustering evolution and has  been often
parametrized as

$$ D^2(z)\ = \ {\left(1+z\right)}^{-(3+\epsilon )}\eqno (8)$$

The case $\epsilon$=$-3$ corresponds to the very peculiar situation of
$r_0$ constant in physical coordinates,
while
$\epsilon =\gamma -3$ implies $r_0$ constant in comoving coordinates.
The case of linear growth of the clustering is represented by
$\epsilon =\gamma -1$ if $\Omega =1$ (Mart{\'\i}nez-Gonzales \& Sanz 1991).
A self-similar evolution of the correlation function (requiring
$\Omega =1$ and a power-law spectrum of initial density perturbations)
yields $\epsilon =0$. In the latter case, on  scales where
$\xi \gg 1$ the number of object in a physical volume is constant
(statistically stable clustering).
Galaxy formation models
like the ``pancake'' model and biased CDM models predict enhanced clustering
of high redshift and luminous QSOs (Rees 1986; Kaiser 1986;
Efstathiou \& Rees 1988).

The redshift dependence of the ACF is determined by the angular scale,
the clustering evolution and the emissivity of the {\it unresolved} sources.
Figure 1 and 2 emphasize the different redshift dependence of the emissivity,
$j_{eff}$, and of the ACF, $W(\theta )$, for some relevant angular separations,
clustering scales and evolution.
The emissivities have been computed using local luminosity
functions and cosmological evolutions consistent with the available
soft and hard X--ray data (see below and Franceschini et al. 1992).
It is apparent that different
combinations of clustering scales and angular separations weight
differently the volume emissivity. For instance, large angular scales,
typical of hard X--ray collimators ($\theta > 2^{\circ }$),
tend to elicit the clustering properties of local sources.
Panels 1b) and 1c) show that for angular separations of $\gsim 2^\circ$
the effect of clustering evolution is important, even
for cases in which the physical clustering scale is significantly decreasing
with increasing redshift, e.g. $\epsilon \geq -1.2$. On the other hand
the effect is negligible in the cases of separations larger than
$4^\circ$ and with $r_0\leq 20$ Mpc, because only nearby objects contribute
to the ACF (panel 1a).

In the soft bands, X--ray telescopes can explore small scales
(few arcminutes) and detect sources down to faint
limiting fluxes; for instance, in deep ROSAT fields up to 50$\%$
of the SXRB has already been resolved (Hasinger 1992).
Small angular separations
make the ACF more sensitive to the clustering of high redshift sources, as
can be seen by comparing Figure 1 with Figure 2. In particular, the cases
reported in panels $c)$ and $d)$ of Figure 2 have been computed
with redshift dependences of the volume emissivity very similar to those used
in Figure 1.
Figure 2 shows that on arcminute scales, the current resolution
achieved by soft bands telescopes, the redshift dependences of the ACF
and of the volume emissivity are very similar. The different relationship
between the ACF passing from degree to arcminute scales,
implies that while the observed HXRB ACF feels only the volume emissivity of
{\it local} sources, that of
the SXRB comes from the integrated contribution to the XRB
of sources at intermediate and high redshift.

Thus, evolution models for the volume emissivity of clustered sources yielding
the same fraction of the background but with a different redshift dependence
could produce rather different values of the ACF at some angular scales.

The dependence of the ACF on  the angular scale
$\theta $, on the clustering scale
$r_{0}$ and on the fraction $f$ of the residual
background produced by the unresolved sources is rather simple
in the case that the beam angular separation $\theta $
and the FWHM of the beam $\theta_{beam}$ are much smaller than the minimum
angle $\theta(r_{0},z)$ subtended by the physical clustering scale.
Then the innermost integration in equation (3) can be extended
from $-\infty $ to $\infty $  and we get (De Zotti et al. 1990, eq. 20):
$$W(\theta )\propto
\theta ^{1-\gamma}  {r_0}^{\gamma } f^{2}. \eqno(9)$$
This relationship holds in the case of ACF studies on small
angular scales; presently this only is the case
of soft X--ray surveys, however in the next future ACF on small
angular scales should also be available at harder energies from
observations performed with ASTRO--D.

So far the ACF derived from available observations in hard
X--ray bands, referring to large angular separations
($\theta  \geq 2^{\circ }$), mostly reflects (particularly in the case
of a steep clustering evolution: $\epsilon \approx 0$) the
clustering properties of rather local ($z \ll 1$) sources which give a small
contribution to the background. In this case, the scaling law is similar
to equation (9), but with $f$ replaced by the ratio of the local volume
emissivity of sources to that of the XRB, $j_{\rm sources}/j_{\rm XRB}$.

\oneskip
\oneskip
\centerline {3. CONSTRAINTS ON X--RAY SOURCE CLUSTERING AND EMISSIVITY}
\centerline{FROM THE SXRB ACF}\nobreak
\oneskip\nobreak
Hasinger, Schmidt \& Tr\"umper (1991) and Hasinger (1992 and private
com\-mu\-ni\-ca\-tion) have derived from deep ROSAT survey data tight $3\sigma$
upper limits to the ACF on arcmin scales:
$W(2'-3')\leq 8\times 10^{-3}$ and $W(9'-10')\leq 2\times 10^{-3}$,
in the 0.9--2.4 keV band, after subtracting sources brighter than
$S_l \approx 5\times 10^{-15}\,\hbox{erg}\,\,\hbox{cm}^{-2}\,\hbox{s}^{-1}$.
Comparable upper limits on the same scales have been found
by preliminary analyses of ROSAT surveys on different areas
by Carrera \& Barcons (1992) and by Georgantopoulos et al. (1992).
Analyzing deep {\it Einstein Observatory} IPC fields
in the energy band 0.8--3.5 keV, Soltan (1991)
derived an even tighter upper limit, $W(2'-5')\leq 3\times 10^{-3}$,
excluding sources brighter than $S_l \approx 5\times 10^{-14}
\,\hbox{erg}\,\hbox{cm}^{-2}\,\hbox{s}^{-1}$;
this limit, however, may be too optimistic (see Barcons \& Fabian 1989
and Carrera \& Barcons 1992).

As shown by the ROSAT deep survey in the QSF3 field (Shanks et al. 1991) the
large majority of the X--ray sources brighter than
$S_l$(0.5--2 keV)$\approx 1\times 10^{-14}\,\hbox{erg}\,\,\hbox{cm}^{-2}
\,\hbox{s}^{-1}$ are
quasars at substantial redshifts. As a consequence AGNs are expected to
give an important contribution to the source counts even below this
limiting flux,
although other source populations like Active Star Forming (ASF) galaxies
could start to appear. Since
it is reasonable to assume that AGNs and ASF galaxies may significantly
contribute to the residual SXRB, the following discussion will be focused
on their clustering and emissivity properties. Although galaxy clusters are
practically absent in the deep surveys, there are models predicting
that clusters could give from 10 to 20$\%$ of the SXRB (Blanchard et al.
1992; Cen \& Ostriker 1992; Cavaliere, Menci \& Burg 1992).
Note that, since these are expected to be
extended, low redshift and low brightness
objects, their contribution to the
autocorrelations on few arcminute scales could not be  negligible, as
in the case of the NEP 'blotch' (Hasinger, Schmidt \& Tr\"umper 1991).

\oneskip
\centerline {\it 3.1. AGNs }\nobreak
\oneskip\nobreak
In Figure 3 we have reported the constraints on the
fraction of the total background contributed by unresolved AGNs and on the
corresponding clustering scale, $r_{0}$, obtained adopting the
ROSAT $3\sigma $ upper limits on the ACF for $9'$--$10'$ scales.
Similar -- albeit somewhat less
stringent -- results can be found comparing model predictions with
limits on $2'$--$4'$ scales.

To avoid a too large contamination by the galactic background, the analysis
of the autocorrelations has been limited to data in the energy range
$0.9\leq E\leq 2.4$ keV, though ROSAT is sensitive down to softer energies
(Hasinger 1992). Henceforth by SXRB we mean the X--ray background
in the energy range $1\lsim E\lsim 3$ keV.
The background intensity in the 0.9-2.4 keV band has been assumed to be
$2\times 10^{-8}\,\hbox{erg}\,\,\hbox{cm}^{-2}\,\hbox{s}^{-1}
\,\hbox{sr}^{-1}$, following Hasinger (1992)
and consistent with McCammon \& Sanders (1990).
{}From table 1 of Hasinger (1992) we have also deduced that
sources brighter than $S\approx 5\times 10^{-15}\,\hbox{erg}\,\,\hbox{cm}^{-2}
\,\hbox{s}^{-1}$ give about $40\%$
of the total background and about $25\%$ of it is attributable to
AGNs (Shanks et al. 1991).

The calculations have been done assuming for the sources a power
law spectrum ($F\propto E^{-\alpha}$)
with spectral index $\alpha =1.2$ as suggested by direct
measurements of bright soft X--ray selected sources and
by the average source spectra observed by ROSAT. We have also
used the Local Luminosity Function derived by Maccaccaro et al. (1991)
and evolution models that fit the EMSS (Gioia et al. 1990)
and ROSAT Deep surveys counts (Hasinger et al. 1991, Shanks et al. 1991,
Anderson et al. 1992, Franceschini et al. 1992). With these assumptions
the fraction of the SXRB contributed by unresolved AGNs is somewhat
dependent on the assumed spectra, luminosity
functions and evolution. Therefore we have decided to keep this fraction
as a free parameter, subject only to the condition that the corresponding
redshift dependence of the emissivity of undetected AGNs is a smooth
extrapolation of that fitting the available data.
The conclusions also depend on the
geometry of the Universe in the sense that changing from
$q_{0}=0.5$ to $q_{0}=0$ the allowed
total fraction of the contributed background increases
by about $\sim 30-40\%$.

It is easy to see from Figure 3 that weak clustering evolution
($\epsilon \leq -1.2$) would produce rather large
autocorrelations even with a small clustering scale $r_{0}$.
As a consequence, if AGNs cluster like galaxies, $r_{0}\simeq 12$
Mpc, the total AGN contribution to the  SXRB is bound to be less than
$\sim 50\%$ ($\sim 25\%$ from detected plus $\sim 25\%$ from unresolved
sources)
in the case $\epsilon = -1.2$ and less than $\sim 35\%$ in the case
$\epsilon =-3$. Because any reasonable model that fits ROSAT deep
counts and EMSS counts predicts a total AGN contribution at least as large as
$30-35\%$ and since a minimum value $r_{0}=12$ Mpc has been found
for optically selected QSOs (Boyle 1991; Andreani \& Cristiani 1992),
clustering evolution must be {\it faster} than that implied by
$\epsilon =-3$ (clustering constant in physical coordinates).

ROSAT results also provide a significant test of the suggestion by
Bahcall \& Chokshi (1991) that the QSO correlation function is
intermediate ($r_{0}=24$ Mpc)
between those of galaxies and of rich clusters, as a consequence of
the preferential location of QSOs in small groups of galaxies,
and has an evolution parameter $\epsilon =-1.8$. From Figure 3
it is apparent that QSOs clustered in this way must produce
less than $40\%$ of the total SXRB (QSOs below the detection
limit yielding less than 15\%).

On the other hand the allowed AGN contribution increases up to
80$\%$ in the case of stable clustering ($\epsilon = 0$).

It is worth noticing that AGNs with clustering scale
$r_{0}=20$ Mpc and $\epsilon = -1.2$, values  still allowed by optical
data, are quite close  to infringe ROSAT upper limits
on the ACF, if they produce $40\%$ of the SXRB.

For sake of completeness
we reported in Figure 4 the constraints derived by using
Soltan's (1991) limits on the ACF, although they could
be too optimistic (Carrera \& Barcons 1992).
At the limiting flux $S_l \approx 5\times
10^{-14}\,\hbox{erg}\,\,\hbox{cm}^{-2}
\,\hbox{s}^{-1}$ (0.8--3.5 keV)
of the surveyed fields AGNs produce only $\sim 20\%$ of the background
(Primini et al. 1991).

\oneskip
\centerline {\it 3.2. Active Star Forming galaxies}\nobreak
\oneskip\nobreak
Several authors (Danese et al. 1987; Hacking, Condon \& Houck 1987;
Franceschini et al. 1988) have pointed out that
Active Star Forming (ASF) galaxies
(galaxies with enhanced
star formation rate in comparison with normal
spiral galaxies) exhibit significant cosmological evolution.
Intense star forming activity should show up also
through X--ray emission powered by supernovae and X--ray binaries
(Bookbinder et al. 1980; Stewart et al. 1982; Griffiths \&
Padovani 1990). Current estimates of the contribution of
the ASF galaxies to the XRB are still rather uncertain
ranging from 5-10$\%$ to 30$\%$ (but an even larger contribution
cannot be excluded), mainly because their X--ray
emission is poorly known. On the other hand
ROSAT deep surveys have shown that QSOs are by far the dominant population
down to the faintest observed fluxes
and only a few of the detected sources could possibly
be identified with ASF galaxies.

We have computed the contribution to the
ACF of ASF galaxies using a luminosity function and luminosity
evolution ($L(z)=L(z=0)\times (1+z)^{2.9}$)
appropriate to keep the population basically
undetected at the ROSAT deep survey limit and to produce
an important fraction of the background. A spectral
index $\alpha=1.5$ has been adopted (see Boller et al. 1992).
The  cumulative fractional emissivity per unit volume as a function
of redshift is shown in panels $c)$ and $d)$ of Figure 2.
The constraints imposed by the smoothness of the ACF on the ASF
clustering are presented in Figure 5.
If we assume a clustering scale $r_{0}\approx 12$ Mpc for these galaxies,
the limits from ROSAT and from  {\it Einstein Observatory}
deep surveys suggest that clustering evolution keeping $r_{0}$ constant in
physical coordinates would entail that ASF galaxies contribute
less than 20$\%$ to the SXRB; the contribution
could reach 40$\%$ in the case of substantial clustering evolution
($\epsilon =0$).
As already suggested by Danese, De Zotti \& Andreani
(1992) the constraints on ASF galaxies could be alleviated only if
one extends to the ASF galaxies the weak clustering found by Efstathiou
et al. (1991) for the faint blue galaxies detected in deep CCD
images (Tyson 1988).

\oneskip
\oneskip
\centerline {4. CONSTRAINTS ON X--RAY SOURCE CLUSTERING AND EMISSIVITY}
\centerline {FROM THE HXRB ACF}\nobreak
\oneskip\nobreak
The autocorrelation function of the HXRB on scales larger than $3^\circ$
has been derived exploiting
data obtained by the HEAO~1 A-2 experiment in the 2--10 keV band
by Persic et al. (1989). More recently, using the same data base
Mart\'{\i}n-Mirones et al. (1991) extended the analysis to scales
smaller than $3^\circ$. Similar analyses have been done by
Carrera et al. (1991, 1992) who have used large sky areas scanned
by the $Ginga$ Large Area Counter in the 4-12 keV band. It is encouraging
that the analyses of both groups produce similar 3$\sigma$ upper limits
on angular separations larger than $2^\circ$, for instance
$W(2^{\circ })\leq 7\times 10^{-4}$.

Mushotzky \& Jahoda (1992), analysing data from the HEAO~1 A-2 experiment on
an area much larger than that exploited by
Persic et al. (1989), have claimed the first detection (99$\%$
confidence level) of autocorrelations in the HXRB with
$W(\theta )\simeq 3\times 10^{-5}$ on angular scales ranging from
6 to 20 degrees. The upper limits found by other authors are
consistent with this possible detection. All the above mentioned analyses
have been done with limiting flux
$S_l \approx 2.5\times 10^{-11}$ $erg\ s^{-1}cm^{-2}$ (2--10 keV)
for source detection.

The distinction we have made between hard and soft XRB, although
arbitrary to some extent, is motivated by some
evidence that the SXRB is mainly contributed by sources like
optically selected QSOs with steep and unabsorbed spectra
(henceforth soft spectrum AGNs) and possibly by ASF galaxies, whereas
the HXRB is mainly contributed by relatively low luminosity AGNs with
heavily absorbed spectra (henceforth hard spectrum AGNs).
The objects dominating the SXRB may give a minor fraction ($f\leq 20\%$)
of the 2--10 keV background, whereas the low luminosity, highly absorbed
objects with hard spectra give a negligible contribution to the
SXRB (see, for a comprehensive
discussion,  Franceschini et al. 1992).

As it is well known, the limits on clustering of optically
selected QSOs refer to high luminosity, high redshift objects (see Boyle
1991; Andreani \& Cristiani 1992). Such limits are thus
more likely to apply to AGNs contributing to the SXRB, while
the clustering properties of low luminosity, heavily
absorbed AGNs could be rather different.

In the following we will use data and limits on the HXRB ACF to explore
clustering and X--ray emission properties of the hard spectrum
AGNs, ASF galaxies and normal galaxies.

\oneskip
\centerline {\it 4.1. Active Galactic Nuclei}\nobreak
\oneskip\nobreak
The constraints ensuing from the limits on the HXRB ACF
at angular scales  smaller than few degrees
have been discussed by Mart\'{\i}n-Mirones et al. (1991), and
Carrera et al. (1991, 1992). The main conclusion
is that the smoothness of the HXRB implies
a rather small clustering scale for the AGNs, if they are required to produce
a significant fraction of the background. We confirm this conclusion.
In particular, using the Piccinotti et al.(1982) local Luminosity Function and
luminosity evolution $L(z)=L(z=0)\times (1+z)^{2.6}$ a fraction
$\sim 60\%$ of the total background  is produced; the
corresponding redshift dependence
of the volume emissivity is presented in Figure 1. In such a case
we have found that the limits on angular scales $\theta \leq 3^{\circ}$
imply $r_0 \leq 25$ Mpc in the case of stable clustering (i.e. $\epsilon =0$)
and, $r_0 \leq 20$ Mpc for a constant comoving clustering scale.

On the other hand  the possible detection of autocorrelations on
a $6^\circ$ scale by Mushotzky
\& Jahoda (1992) would imply $r_0 \simeq 20$ Mpc,
independently of the clustering evolution and of the geometry
of the Universe, because the contribution
to the $6^\circ$ ACF of objects clustered on
linear scales $r_0 \leq 20$ Mpc is confined to low redshifts,
as can be seen in Figure 1a. Actually $\sim 80\%$ of the total
autocorrelation is produced by AGNs within
$z\lsim 0.07$ giving only $\sim 1\%$
of the total background, based on the local luminosity function by
Piccinotti et al. (1982).

Clustering evolution keeping constant the physical scale ($\epsilon =-3$)
does not exclude the possibility of a relevant contribution (about 60$\%$)
of hard spectrum AGNs to the HXRB, provided that $r_0 \leq 12$ Mpc.

The limits imposed by the HXRB autocorrelations are looser than those
derived using the soft band data.
Apart from the fact that ROSAT ACF studies (Hasinger 1992)
refer to a residual 60$\%$ of the background, the tightness of the
limits imposed by the SXRB ACF is due to the imaging capability of the
telescopes that allows to explore the ACF on small angular scales.
Actually, as discused in Section 2, only a small fraction
of the total emissivity of the population contributes to the
autocorrelations in the hard band case, because large angular scales
emphasize low redshift contributions when the explored linear clustering
scales are of the order of few tens of megaparsecs.
Contrariwise, in the soft band the fact that the ACF can be studied on
small angular scales (from few to several arcminutes)
ensures that $\theta \ll \theta(r_{0},z)$ at any redshifts
and, as a consequence, that the overall source population contributes
to the autocorrelations. On the other hand this fact makes the soft
and hard data on the ACF complementary, because they are sensitive mainly
to high and low redshift source clustering respectively.

\oneskip
\centerline {\it 4.2. Active Star Forming and normal galaxies}\nobreak
\oneskip\nobreak

If we assume that the typical clustering scale of ASF galaxies
is similar to that of normal galaxies ($r_0 \approx 10-12$ Mpc),
the limits on the HXRB ACF do not set any significant constraint on
their contribution to the XRB, even in the case of a constant physical
clustering scale, due to their low local volume emissivity.
For example, we have assumed that ASF galaxies have
a present number density
$n_{\rm ASF}\approx 1.5\times 10^{-3}\,\hbox{Mpc}^{-3}$,
close to that of bright galaxies, and an average
2--10 keV luminosity $\langle L_x\rangle \approx 1 \times 10^{41}
\,\hbox{erg}\,\hbox{s}^{-1}$.
The local volume emissivity is then $\rho_x (2\hbox{--}10\,\hbox{keV})\approx
1.5\times 10^{38}\,\hbox{erg}\,\hbox{s}^{-1}\,\hbox{Mpc}^{-3}$, a factor of
5 higher than
the limit found by Rephaeli et al. (1991);
with a luminosity evolution
$L(z)=L(z=0)\times (1+z)^{3.4}$ about $60\%$ of the
HXRB is produced.

Even under these extreme assumptions, this source population
yields autocorrelations $W(2^{\circ })\leq 2\times 10^{-5}$,
much smaller than the observational upper limits even
in the case $\epsilon=-3.$
Note however that the case for a dominant contribution to the HXRB
from the ASF galaxies
is unlikely, because their observed  spectra
(see e.g. Boller et al. 1992) are much steeper than those, flat or even
inverted, required to account for the HXRB after removal of contributions
of the already detected sources (see Franceschini et al. 1992).

As shown in \S 2, the contribution to the ACF on a $6^\circ$ scale
depends on the local volume emissivity. In particular with a clustering
scale $r_0=12$ Mpc, typical of galaxies, it can be seen that

$$W(6^{\circ })\approx 1-0.8\times 10^{-6}\left({\rho_x\over 10^{38}}
\right )^2, \eqno (10)$$
depending on the redshift the bulk of the HXRB is assumed to come from.

We conclude that ASF galaxies cannot
significantly contribute to the autocorrelations
reported by Mushotzky \& Jahoda (1992), even in the case they eventually
produce a dominant fraction of the HXRB.

Cross-correlations of the XRB surface brightness
with the projected distribution of galaxies, both optically and far-IR
selected, have
been used to evaluate or constrain the X--ray volume emissivity
associated with galaxies (Turner \& Geller 1980; Jahoda et al.
1991; Lahav 1992; Boldt 1992; Jahoda et al. 1992). In particular
Jahoda et al. (1992)
using HEAO~1 A-2 data and the UGC and ESO catalogs have been able to
estimate the 2--10 keV X--ray luminosity density of the local Universe
$\rho_x (2-10 keV)=
(12.5\pm 7)\times 10^{38}$ $erg \ s^{-1} Mpc^{-3}$.
Relating the peculiar velocity of the
local group of galaxies to the total dipole moment of the all--sky
distribution of the X--ray flux, the same authors have estimated
the present epoch volume emissivity
$\rho_x (2-10\ keV)\sim 30
\times 10^{38} (b\Omega^{-0.6})^{-1}$ $erg \ s^{-1} Mpc^{-3}$ (b is
the bias parameter for low luminosity X--ray sources). Boldt (1992)
has also estimated the 2--10 keV local volume emissivity of normal
galaxies $\rho_x (2-10\ keV)\approx
3 \times 10^{37}$ $erg \ s^{-1} Mpc^{-3}$,
extrapolating the results found by Fabbiano, Kim \& Trinchieri (1992)
in the soft bands.
The insertion of the latter value of the volume emissivity in equation (10)
shows that the expected contribution from normal galaxies to the ACF
at $6^\circ$ is negligible.

On the other hand, after equation (10), the limits on the HXRB ACF entail that
the local volume emissivity of low luminosity sources
clustered like normal galaxies is bound to be $\rho_x
(2-10\ keV)\lsim 6\times 10^{38}\,\hbox{erg}\,\hbox{s}^{-1}\,\hbox{Mpc}^{-3}$.
This value is half of that derived from the cross--correlation of the
background with nearby galaxies and even lower than the estimate obtained
through the XRB dipole (Jahoda et al. 1992). However the errors associated
to Jahoda et al. (1992) estimates of local volume emissivity are so
large that they are consistent at 1$\sigma$ with our limit. Moreover
as pointed out by Jahoda (private communication) their estimate could be
affected both by uncertainties in modelling the relationship between
fluctuations
in galaxy counts and XRB intensity as well as by the overdensity of the
local universe.

\oneskip
\oneskip
\centerline {5. DISCUSSION AND CONCLUSIONS}\nobreak
\oneskip\nobreak
The smoothness of the SXRB significantly constrains the
clustering scale, the clustering evolution and the contribution to the
background of soft spectrum AGNs. The limits discussed in Section
3.1 can be straightforwardly compared with the results derived from optical
surveys of QSOs (Iovino, Shaver \& Cristiani 1991; Boyle 1991; Andreani
\& Cristiani 1992), because there is evidence that soft X--ray
and optical selections tend to single out the same class of AGNs
(Shanks et al. 1991; Setti 1991).

Recent analyses of large samples of optically selected QSOs
have produced  consistent values of the clustering scale
$12\lsim r_0\lsim 20$ Mpc (Iovino, Shaver \& Cristiani 1991; Boyle 1991;
Andreani
\& Cristiani 1992). The same authors also agree on the fact
that there is evidence of evolution of the correlation function with
$\epsilon \geq -1.2$, a constant
comoving clustering scale being slightly favoured.

The limits derived in Section 3.1 are compatible with the
optical results. It is interesting to notice that  clustering
with $r_0=20$ Mpc and $\epsilon =-1.2$ would imply that
soft spectrum AGNs produce less than 40$\%$ of the SXRB and that
the limit  shifts to 50$\%$ if $r_0=12$ Mpc. Indeed
plausible models that are consistent with a number of observations,
such as the local luminosity function, ROSAT and EINSTEIN Observatory source
counts,
redshift distributions and spectral properties of soft X--ray selected
AGNs, predict an AGN contribution to the SXRB
in the range between 30 to 50$\%$ (Franceschini et al. 1992).
Thus, in the case of AGNs, clustering evolution with $\epsilon =-1.2$ is
just on the verge of producing a too large ACF. Whatever the
origin of  the remaining background is, its sources must cluster rather
weakly.
Galaxy clusters and ASF galaxies are obvious candidates to produce
the remaining $\sim$50$\%$ of the SXRB. Galaxy clusters are known to cluster
on large scale, the values of $r_0$ found by various authors ranging from
$r_0 \approx 50$ to $r_0 \approx 30\,$Mpc  (Bahcall \& Soneira 1983; Postman,
Geller \& Huchra 1986; Sutherland 1988; Sutherland \& Efstathiou 1991;
Bahcall \& West 1992).
As a consequence their contribution to the background is bound to be small
($f\lsim 15\%$), unless they are very extended and of low surface
brightness. On the other hand
ASF galaxies could yield a significant fraction of the SXRB, because
they probably cluster on a scale comparable (possibly smaller, cfr.
Efstathiou et al. 1991) to that of normal galaxies.

It is worth noticing that the constraints on both AGNs and ASF galaxies
have been obtained under the hypothesis that any other contribution
to the background is
smoothly distributed in the sky. Therefore constraints on clustering
are expected to be even more stringent than those derived above.
For instance let us consider the case in which
AGNs and ASF galaxies
produce together a fraction f$\simeq$80$\%$ of the SXRB
(leaving room for contributions
of galaxy clusters and galactic stars). Let us also assume that
AGNs cluster with $r_0=12$ Mpc and $\epsilon= -1.2$ and give
about 50$\%$ of the background; as a consequence they would saturate
the autocorrelation level allowed by ROSAT limits
(cfr. Figure 3). Then ASF galaxies giving
the residual fraction $f\simeq 30\%$ are bound to have $r_0\lsim 10$
Mpc even in the case $\epsilon=0$ [cf. Figure 5 and eq. (9)].

As for the HXRB, Mushotzky \& Jahoda (1992) claimed
a possible detection of positive autocorrelations on large
scales, $\theta \geq 6^{\circ}$. This signal could well be due to
nearby, hard spectrum AGNs clustered with $r_0\approx 20$ Mpc.
Moreover, if the clustering evolution is confined to
$\epsilon \geq -1.2$, AGNs could also produce $\sim 60\%$ of the HXRB,
without violating the presently available limits on
autocorrelations on scales of a few degrees (Mart{\'\i}n-Mirones
et al. 1991; Carrera et al. 1991; 1992). Of course
these AGNs are bound to give
a negligible contribution to the SXRB; this could be the
result of an absorption increasing with increasing redshift
(Franceschini et al. 1992).

As pointed out by Danese, De Zotti \& Andreani (1992) and by
Carrera \& Barcons (1992), such a level of clustering cannot
significantly affect the probability density distribution
of the deflections from the mean, $P(D)$ observed by HEAO~1 A-2
(Shafer 1983) as well as by $Ginga$ (Warwick \& Stewart 1989).

Clustering of galaxy clusters yields
$W(6^{\circ })\lsim 1 \times 10^{-5}$, if
$r_0\approx 30-50$ Mpc, as found in optical
surveys (Bahcall \& West 1992).

An alternative explanation of the possible detection
of large scale autocorrelations
requires that the local volume emissivity of the low luminosity sources is
$\rho_x (2-10 keV)\approx
6\times 10^{38}$ $erg \ s^{-1} Mpc^{-3}$.


Conversely the above
value  can be also interpreted as an upper
limit to the local volume emissivity
leading to the conclusion
that the sources contributing the bulk of the HXRB are not local.
This is somewhat in contrast with the findings by Jahoda et al. (1992)
(see above \S 4) and Lahav et al. (1992), who claim a much larger
local volume emissivity. Apart from the large uncertainties
associated to their estimates, the only possible way to make
their and our results consistent
is that the low luminosity sources responsible for the local emissivity
at the level claimed by these authors are clustered on a typical
scale $r_{\circ}\approx 6$ Mpc, half of the value found for normal
galaxies.

All in all, the results on ACF found by Mushotzky \& Jahoda (1992)
significantly constrains the local volume emissivity.

Important developments in ACF studies
should come out from the analysis of the ASTRO--D
data. Imaging capabilities with resolutions of few arcminutes will be
obtained by this experiment in the 2--10 keV band and sources
brighter than some $10^{-14} \,\hbox{erg}\,\,\hbox{cm}^{-2}\,\hbox{s}^{-1}$
will be detected (Inoue 1992). We have computed the expected ACF in the
hypothesis that the bulk of the HXRB is produced at substantial redshifts
(the cumulative fraction of the effective volume emissivity
has been reported in panel a) of Figure 1) and for a
limiting flux of $S_l\sim 1\times 10^{-13}$. Assuming
$\gamma =1.8$ and following equation (9), we have found
$$W(\theta )\approx g(\epsilon)
\left({\theta \over 5^\prime}\right)  ^{-0.8}
\left({r_0 \over {12\ Mpc}}\right)  ^{1.8}
\left({f \over 0.5}\right)^{2}, \eqno(11)$$
where g($\epsilon $) is a function of the clustering evolution; g($\epsilon $)
is $4\times 10^{-3}$, $1\times 10^{-2}$ and $8\times 10^{-2}$ for
$\epsilon =0$, $\epsilon =-1.2$ and $\epsilon =-3$ respectively.
The availability of ACF data on large and small
angular scales would then allow
to extract important information on the clustering of sources
contributing to the HXRB
both in the local as well as in the high redshift universe.

Finally, it is worth noticing that the limits on clustering
derived from SXRB and HXRB correlations suggest that soft, highly
luminous AGNs and hard, less luminous AGNs have rather similar clustering
properties, with scales not much larger than that of the galaxies,
and evolution at least as steep as implied by the comoving
clustering model, i.e. $\epsilon \geq -1.2$.
\oneskip
\oneskip
\oneskip\noindent
{\it Acknowledgements} We thank X. Barcons, F. Carrera,
G. Hasinger and G. Setti for enlightening discussions.
Comments and suggestions of an anonimous referee and of K. Jahoda
have been extremely helpful.
Work supported in
part by MURST, GNA/CNR and ASI.

\vfill\eject

\oneskip
\oneskip
\oneskip
\centerline {\bf REFERENCES}
\oneskip
\parindent=0pt
\parskip=0pt

\ref
Anderson, S.F., Windhorst, R.A., Maccacaro, T., Burstein, D., Franklin, B.E.,
Griffiths, R.E., Koo, D.C., Mathis, D.F., Morgan, W.A. \& Neuschafer, L.W.
1992, in X--ray Emission from Active Galactic Nuclei and the Cosmic X--ray
Background, eds. W. Brinkmann \& J. Tr\"umper, MPE Report 235, p. 227

\ref
Andreani, P., \& Cristiani, S. 1992, ApJ, 398, L13

\ref
Bahcall, N.A., \& Chokshi, A. 1991, \apj 380 L9/

\ref
Bahcall, N.A., \& Soneira, R.M. 1983, ApJ, 270, 20

\ref
Bahcall, N.A., \& West, M.J. 1992, ApJ, 392, 419

\ref
Barcons, X., \& Fabian, A.C. 1989, MNRAS, 237, 119

\ref
Blanchard, A., Wachter, K., Evrard, A.E., \& Silk, J. 1992,
\apj 391 1/

\ref
Boldt, E.A. 1987, Phys. Rep., 146, 215

\ref
Boldt, E. 1992, in The X--ray Background, eds. X. Barcons \& A.C.
Fabian (Cambridge University Press), 115

\ref
Boller, Th., Meurs, E.J.A., Brinkmann, W., Fink, H., Zimmermann, U., \&
Adorf, H.-M. 1992, in X--ray Emission from Active Galactic Nuclei and the
Cosmic X--ray Background, eds. W. Brinkmann \& J. Tr\"umper, MPE Report 235,
p.231

\ref
Bookbinder, J., Cowie, L.L., Krolik, J.H., Ostriker, J.P., \& Rees, M.
1980, \apj 237 647/

\ref
Boyle, B.J. 1991, in Texas-ESO/CERN Conference on Relativistic
Astrophysics, in press

\ref
Carrera, F.J., \& Barcons, X. 1992, \mnras 257 507/

\ref
Carrera, F.J., Barcons, X., Butcher, J.A., Fabian, A.C., Stewart, G.C.,
Warwick, R.S., Hayashida, K. \& Kii, T. 1991, \mnras 249 698/

\ref
Carrera, F.J., Barcons, X., Butcher, J.A., Fabian, A.C., Stewart, G.C.,
Toffolatti, L., Warwick, R.S., Hayashida, K., Inoue, H., \& Kondo, H. 1992,
MNRAS, in press

\ref
Cavaliere, A., Menci, N., \& Burg, R. 1992, preprint

\ref
Cen, R., \& Ostriker, J. 1992, ApJ, 393, 22

\ref
Danese, L., De Zotti, G., \& Andreani, P. 1992,
in The X--ray Background, eds. X. Barcons \& A.C.
Fabian (Cambridge University Press), 61

\ref
Danese, L., De Zotti, G., Franceschini, A., \& Toffolatti, L. 1987,
\apj 318 L15/

\ref
De Zotti, G., Persic, M., Franceschini, A., Danese, L., Palumbo, G.G.C.,
Boldt, E.A., \& Marshall, F.E. 1990, ApJ, 351, 22

\ref
Efstathiou, G., Burnstein, G., Katz, N., Tyson, J.A., \& Guhathakurta,
P. 1991, \apj 380 L47/

\ref
Efstathiou, G., \& Rees, M.J. 1988, MNRAS, 230, 5P

\ref
Fabbiano, G., Kim, D.W., \& Trinchieri, G. 1992, ApJS, in press

\ref
Fabian, A.C. 1972, Nature Phys. Sci., 237, 19

\ref
Franceschini, A., Danese, L., De Zotti, G., \& Toffolatti, L. 1988,
\mnras 233 157/

\ref
Franceschini, A., Mart\'{\i}n-Mirones, J.M., Danese, L., \& De Zotti, G. 1992,
MNRAS, in press

\ref
Georgantopoulos, I., Stewart, G.C., Griffiths, R.E., Shanks, T., \& Boyle, B.
1992, in X--ray Emission from Active Galactic Nuclei and the Cosmic X--ray
Background, eds. W. Brinkmann \& J. Tr\"umper, MPE Report 235, p. 368

\ref
Gioia, I.M., Henry, J.P., Maccacaro, T., Morris, S.L., Stocke, J.T., \&
Wolter, A. 1990, \apj 356 L35/


\ref
Griffiths, R.E., \& Padovani, P. 1990, \apj 360 483/

\ref
Hacking, P., Condon, J.J., \& Houck, J.R. 1987, \apj 316 L15/

\ref
Hasinger, G. 1992, in The X--ray Background, eds. X. Barcons \& A.C.
Fabian (Cambridge University Press), 229

\ref
Hasinger, G., Schmidt, M., \& Tr\"umper, J. 1991, \aa 246 L2/

\ref
Inoue, H., 1992, in The X--ray Background, eds. X. Barcons \& A.C.
Fabian (Cambridge University Press), 286

\ref
Iovino, A., Shaver, P.A., \& Cristiani, S. 1991, in The Space Distribution
of Quasars, ed. D. Crampton, ASP Conference Series n. 21, p. 202

\ref
Jahoda, K., Lahav, O., Mushotzky, R.F., Boldt, E.A. 1991, \apj 378 L37/

\ref
Jahoda, K., Lahav, O., Mushotzky, R.F., Boldt, E.A. 1992, \apj 399 L107/

\ref
Kaiser, N. 1986, MNRAS, 222, 323

\ref
Lahav, O. 1992, in The X--ray Background, eds. X. Barcons \& A.C.
Fabian (Cambridge University Press), 102

\ref
Lahav, O., Fabian, A.C., Barcons, X., Boldt, E., Butcher, J., Carrera, F.J.,
Jahoda, K., Miyaji, T., Stewart, G.C., \& Warwick, R.S. 1992, Nature, in press

\ref
Maccacaro, T., Della Ceca, R., Gioia, I.M., Morris, S.L., Stocke J.T.,
\& Wolter, A. 1991, \apj 374 117/

\ref
Maddox, S.J., Efstathiou, G., Sutherland, W.J., \& Loveday, J. 1990,
\mnras 242 43p/


\ref
Marshall, F.E., Boldt, E.A., Holt, S.S., Miller, R.B., Mushotzky, R.F.,
Rose, L.A., Rothschild, R.E., \& Serlemitsos, P.J. 1980, \apj 235 4/

\ref
Mart\'{\i}nez-Gonz\'alez, E., \& Sanz, J.L. 1991, MNRAS, 248, 816

\ref
Mart\'{\i}n-Mirones, J.M., De Zotti, G., Boldt, E.A., Marshall, F.E.,
Danese, L., Franceschini, A., \& Persic, M. 1991, \apj 379 507/

\ref
Mather, J.C. et al. 1990, \apj 354 L37/

\ref
McCammon, D., \& Sanders, W.T. 1990, \araa 28 657/


\ref
Micela, G., Harnden, F.R., J.R., Rosner, R., Sciortino, S., \& Vaiana, G.S.
1991, \apj 380 495/

\ref
Mushotzky, R., \& Jahoda, K. 1992,
in The X--ray Background, eds. X. Barcons \& A.C.
Fabian (Cambridge University Press), 80

\ref
Peebles, P.J.E. 1980, The Large-Scale Structure of the Universe
(Princeton: Princeton University Press)

\ref
Persic, M., De Zotti, G., Boldt, E.A., Marshall, F.E., Danese, L.,
Franceschini, A., \& Palumbo, G.G.C. 1989, \apj 336 L47/

\ref
Postman, M., Geller, M.J., \& Huchra, J.P. 1986, \aj 91 1267/

\ref
Primini, F.A., Murray, S.S., Huchra, J., Schild, R., Burg, R., \&
Giacconi, R. 1991, \apj 374 440/

\ref
Rephaeli, Y., Gruber, D., Persic, M., \& MacDonald, D. 1991, \apj 380
L59/

\ref
Rees, M.J. 1986, in The Structure and Evolution of Active Galactic Nuclei,
eds. G. Giuricin, F. Mardirossian, M. Mezzetti, \& M. Ramella (Dordrecht:
Reidel), p. 447

\ref
Schwartz, D.A., Boldt, E.A., Holt, S.S., Serlemitsos, P.J., \& Bleach,
R.D. 1971, Nature Phys. Sci., 233, 110

\ref
Sebok, W.L. 1986, ApJS, 62, 301

\ref
Setti, G. 1991, in Frontiers of X--ray Astronomy, Proceedings of the
28th Yamada Conference, eds. Y. Tanaka \& K. Koyama (Universal Academy
Press), 663

\ref
Shafer, R.A. 1983, Ph. D. thesis, University of Maryland

\ref
Shanks, T., Georgantopoulos, I., Stewart, G.C., Pounds, K.A.,
Boyle, B.J., \& Griffiths, R.E. 1991, \nat 353 315/

\ref
Soltan, A.M. 1991, \mnras 250 241/

\ref
Stewart, G.C., Fabian, A.C., Terlevich, R.J., \& Hazard, C. 1982,
\mnras 200 61p/

\ref
Sutherland, W.J. 1988, in IAU Symposium 130, Large Scale Structures
of the Universe, ed J. Audouze, M.C. Pelletan, and A. Szalay (Dordrect:
Kluwer), p. 538

\ref
Sutherland, W.J., \& Efstathiou, G. 1991, \mnras 248 159/

\ref
Turner, E.L., \& Geller, M.J. 1980, \apj 236 1/

\ref
Tyson, J.A. 1988, \aj 96 1/

\ref
Wolfe, A.M., \& Burbidge, G.R. 1970, \nat 228 1170/

\ref
Wu, X., Hamilton, T., Helfand, D.J., \& Wang, Q. 1991, \apj 379 564/

\vfill\eject

\oneskip
\centerline {\bf FIGURE CAPTIONS}
\oneskip
\parindent=0pt
\parskip=0pt

{\bf Figure 1} Predicted cumulative fraction of the ACF ($W(\theta )$, {\it
continuous line}) and of
the effective volume emissivity ($j_{eff}$, {\it short-dashed line})
per unit logarithmic redshift interval for hard X--ray selected AGNs.
The relevant parameters and the angular scale are indicated in each panel.

\oneskip
{\bf Figure 2} Same as in Figure 1 but for soft X--ray selected sources.
Panels a) and b) refer to AGNs while panels c) and d)
refer to Active Star Forming galaxies (see Section 3.2).

\oneskip
{\bf Figure 3} Constraints on the fraction ($f$) of the total background
contributed by unresolved AGNs as a function of the clustering scale, $r_{0}$.
The plotted curves, following the scaling law of equation (9), are obtained
by saturating the $3\sigma$ upper limit on $W(9^{\prime})$ from ROSAT deep
exposures (Hasinger 1992). The different curves refer to different values
of the clustering evolution parameter $\epsilon$ as indicated by the labels.
The current unresolved fraction of the SXRB is indicated by the
{\it dotted-long-dashed} line ($f=0.6$) (see text for details).



\oneskip
{\bf Figure 4} Same as in Figure 3 but using the $3^{\prime}$ upper limit
obtained by Soltan (1991) from the analysis of {\it Einstein} IPC data.
The {\it dotted-long-dashed} line again indicate the fraction of the
unresolved background ($f=0.8$) at the limit of the {\it Einstein} Deep Survey.

\oneskip
{\bf Figure 5} Same as in Figure 3 but for ASF galaxies.


\vfill\eject
\end

\ref
Bagoly, Z., M\'esz\'aros, A., and M\'esz\'aros, P. 1988, ApJ, 333, 54

\ref
Bahcall, N.A., Batuski, D.J., and Olowin, R.P. 1988, ApJ (Letters),
333, L13

\ref
Barcons, X., and Fabian, A.C. 1988, MNRAS, 230, 189

\ref
Batuski, D.J., Bahcall, N.A., Olowin, R.P., and Burns, J.O. 1989, ApJ,
341, 599

\ref
Carrera, F.J., Barcons, X., Butcher, J., Fabian, A.C., Stewart, G.C.,
Warwick, R.S., Hayashida, K., and Kii, T. 1991, MNRAS, in press

\ref
Cavaliere, A., and Colafrancesco, S. 1988, ApJ, 331, 660

\ref
Cavaliere, A., and Colafrancesco, S. 1989, in {\it Large Scale Structure and
Motions in the Universe}, ed. M. Mezzetti et al. (Dordrecht: Kluwer), p. 73.

\ref
Cavaliere, A., Giallongo, E., and Vagnetti, F. 1985, ApJ, 296, 402

\ref
Cole, S., and Kaiser, N. 1988, MNRAS, 233, 637

\ref
Danese, L., De Zotti, G., Fasano, G., and Franceschini, A. 1986, A{\&}A,
161, 1

\ref
Danese, L., and Franceschini, A. 1988, in {\it The Post--Recombination
Universe}, ed. N. Kaiser and A.N. Lasenby (Dordrecht: Kluwer), p. 33

\ref
Dautcourt, G. 1977, Astr. Nachr., 298, 141

\ref
De Zotti, G., Persic, M., Franceschini, A., Danese, L., Palumbo, G.G.C.,
Boldt, E.A., and Marshall, F.E. 1989, in {\it IAU Symp. No. 134: Active
Galactic Nuclei}, ed. D.E. Osterbrock and J.S. Miller (Dordrecht: Kluwer),
p. 492

\ref
Fabian, A.C., George, I.M., Miyoshi, S., and Rees, M.J. 1990, MNRAS,
242, 14P

\ref
Giacconi, R., et al. 1979, ApJ (Letters), 234, L1

\ref
Goicoechea, L.J., and Mart\'{\i}n-Mirones, J.M. 1990, MNRAS, 244, 493

\ref
Iovino, A., et al. 1989, in {\it Large Scale Structure and Motions in the
Universe}, ed. M. Mezzetti et al. (Dordrecht: Kluwer), p. 369

\ref
Jahoda, K., and Mushotzky, R.F. 1989, ApJ, 346, 638

\ref
Klypin, A.A., and Kopylov, A.I. 1983, Pis'ma A. Zh., 9, 75 [Sov. Astr.
Letters, 9, 41]

\ref
Marshall, F.E., Boldt, E.A., Holt, S.S., Mushotzky, R.F., Pravdo, S.H.,
Rothschild, R.E., and Serlemitsos, P.J. 1979, ApJS, 40, 657

\ref
Matsuoka, M., Piro, L., Yamauchi, M., and Murakami, T. 1990, ApJ, 361, 440

\ref
M\'esz\'aros, A., and M\'esz\'aros, P. 1988, ApJ, 327, 25

\ref
Miyaji, T., and Boldt, E. 1990, ApJ (Letters), 353, L3

\ref
Morisawa, K., Matsuoka, M., Takahara, F., and Piro, L. 1990, A\& A, 236, 299

\ref
Mushotzky, R.F. 1988, in {\it Hot Thin Plasmas in Astrophysics}, ed. R.
Pallavicini (Dordrecht: Reidel), p. 273

\ref
Mushotzky, R.F. 1989, {\it Proc. 23rd ESLAB Symposium,
Vol. 2, ``AGN and the X--ray Background''}, ed. J. Hunt and B. Battrick,
ESA SP-296, p. 857

\ref
Olivier, S., Blumenthal, G.R., Dekel, A., Primack, J.R., and Stanhill,
D. 1990, ApJ, 356, 1

\ref
Piccinotti, G., Mushotzky, R.F., Boldt, E.A., Holt, S.S., Marshall, F.E.,
Serlemitsos, P.J., and Shafer, R.A. 1982, ApJ, 253, 485

\ref
Rothschild, R.E., et al. 1979, Space Sci. Instr., 4, 269

\ref
Schmidt, M., Hasinger, G., and Tr\"umper, J. 1991, in
{\it Proc. XIth Moriond Astrophysics Meeting}, in press

\ref
Setti, G. 1987, in {\it IAU Symposium 124, Observational Cosmology}, ed.
A. Hewitt, G. Burbidge, and L.Z. Fang (Dordrecht:Reidel), p. 579

\ref
Setti, G., and Woltjer, L. 1989, A\& A, 224, L21

\ref
Shafer, R.A., and Fabian, A.C. 1983, in {\it IAU Symposium 104, Early
Evolution of the Universe and its Present Structure}, ed. G.O. Abell and G.
Chincarini (Dordrecht:Reidel), p. 333

\ref
Shaver, P.A., Iovino, A., and Pierre, M. 1989, in {\it Large Scale
Structure and Motions in the Universe}, ed. M. Mezzetti et al.
(Dordrecht: Kluwer), p. 101

\ref
Sutherland, W.J. 1988, in {\it IAU Symposium 130, Large Scale Structures of
the Universe}, ed. J. Audouze, M.-C. Pelletan, and A. Szalay (Dordrecht:
Kluwer), p. 538

\ref
Totsuji, H., and Kihara, T. 1969, PASJ, 21, 221

\ref
Warwick, R.S., and Stewart, G.C. 1989, {\it Proc. 23rd ESLAB Symposium,
Vol. 2, ``AGN and the X--ray Background''}, ed. J. Hunt and B. Battrick,
ESA SP-296, p. 727

\ref
Weinberg, S. 1972, {\it Gravitation and Cosmology} (New York:Wiley)

\vfill\eject

\ref
Abell, G.O. 1958, {\it Ap. J. Suppl.}, {\bf 3}, 211.

\ref
Bahcall, N.A., and Burgett, W. 1986, {\it Ap. J. (Letters)}, {\bf 300}, L15.

\ref
Barnes, J., Dekel, A., Efstathiou, G., and Frenk, C.S. 1985, {\it Ap. J.},
{\bf 295}, 368.

\ref
Clowes, R.G. 1986, {\it M.N.R.A.S.}, {\bf 218}, 139.

\ref
Clowes, R.G., Iovino, A., and Shaver, P. 1987, {\it M.N.R.A.S.}, {\bf 227},
921.

\ref
Cole, S., and Kaiser, N. 1988, {\it M.N.R.A.S.}, {\bf 233}, 637.

\ref
Crampton, D., Cowley, A.P., and Hartwick, F.D.A. 1988, in {\it Proceedings
of a Workshop on Optical Surveys for Quasars}, ed. P.S. Osmer, A.C. Porter,
R.F. Green, and C.B. Foltz (San Francisco:Astronomical Society of the
Pacific), p. 254.

\ref
Danese, L., and De Zotti, G. 1986, in {\it Galaxy Distances and Deviations
from Universal Expansion}, ed. B.F. Madore and R.B. Tully (Dordrecht:Reidel),
p. 215.

%
\ref
De Zotti, G., Persic, M., Franceschini, A., Danese, L., Palumbo, G.G.C.,
Boldt, E.A., and Marshall, F.E. 1989, in {\it IAU Symp. No. 134: Active
Galactic Nuclei}, ed. D.E. Osterbrock and J.S. Miller (Dordrecht: Kluwer),
p. 492.

\ref
De Zotti, G., Persic, M., Franceschini, A., Danese, L., Palumbo, G.G.C.,
Boldt, \& E.A., Marshall, F.E. 1990, ApJ, 351, 22

\ref
Drinkwater, M.J. 1988, {\it M.N.R.A.S.}, {\bf 235}, 1111.

\ref
Efstathiou, G., and Rees, M.J. 1988, {\it M.N.R.A.S.}, {\bf 230}, 5P.

\ref
Einasto, J., Klypin, A.A., and Saar, E. 1986, {\it M.N.R.A.S.}, {\bf 219}, 457.

\ref
Fabian, A.C 1981, in {\it Proc. Tenth Texas Symposium on Relativistic
Astrophysics}, {\it Ann. N.Y. Acad. Sci.}, {\bf 375}, 235.

\ref
Fabian, A.C. 1988, in {\it The Post--Recombination Universe}, ed. N. Kaiser
and A.N. Lasenby (Dordrecht: Kluwer), p. 51.

\ref
Huchra, J., Geller, M., de Lapparent, V., and Burg, R. 1988, in
{\it IAU Symposium 130, Large Scale Structures of
the Universe}, ed. J. Audouze, M.-C. Pelletan, and A. Szalay (Dordrecht:
Kluwer), p. 105.

\ref
Johnson, M.W., Cruddace, R.G., Ulmer, M.P., Kowalski, M.P., and Wood, K.S.
1983, {\it Ap. J.}, {\bf 266}, 425.

%

\ref
Kowalski, M.P., Ulmer, M.P., and Cruddace, R.G. 1983, {\it Ap. J.}, {\bf 268},
540.

\ref
Kruszewski, A. 1989, in {it Large Scale Structure and Motions in the Universe},
ed. M. Mezzetti, G. Giuricin, F. Mardirossian, and M. Ramella (Dordrecht:
Kluwer), p. 385.

\ref
Piro, L., Yamauchi, M., and Matsuoka, M. 1990, ApJ (Letters), in press

\ref
Mushotzky, R.F. 1982, {\it Ap. J.}, {\bf 256}, 92.

\ref
Olowin, R.P., Chincarini, G., and Corwin, H.G. 1987, {\it Bull. AAS},
{\bf 19}, 1073.

\ref
Persic, M., Rephaeli, Y., and Boldt, E.A. 1988, {\it Ap. J. (Letters)},
{\bf 327}, L1.

\ref
Rees, M.J. 1986, in {it The Structure and Evolution of Active Galactic Nuclei},
ed. G. Giuricin, F. Mardirossian, M. Mezzetti, and M. Ramella (Dordrecht:
Reidel), p. 447.

\ref
Rothschild, R.E., Mushotzky, R.F., Baity, W.A., Gruber, D.E., Matteson,
J.L., and Peterson, L.E. 1983, {\it Ap. J.}, {\bf 269}, 423.

\ref
Schechter, P. 1976, {\it Ap. J.}, {\bf 203}, 297.

\ref
Shanks, T., Boyle, B.J., and Peterson, B.A. 1988, in {\it Proceedings
of a Workshop on Optical Surveys for Quasars}, ed. P.S. Osmer, A.C. Porter,
R.F. Green, and C.B. Foltz (San Francisco:Astronomical Society of the
Pacific), p. 244.

\ref
Shanks, T., Fong, R., Boyle, B.J., and Peterson, B.A. 1987, {\it M.N.R.A.S.},
{\bf 227}, 739.

\ref
Shaver, P.A. 1984, {\it Astr. Ap.}, {\bf 136}, L9.

\ref
Shaver, P.A. 1988, in {\it IAU Symposium 130, Large Scale Structures of
the Universe}, ed. J. Audouze, M.-C. Pelletan, and A. Szalay (Dordrecht:
Kluwer), p. 359.

\ref
Shectman, S.A. 1985, {\it Ap. J. Suppl.}, {\bf 57}, 77.

\ref
White, S.D.M., Frenk, C.S., Davis, M., and Efstathiou, G. 1987, {\it Ap. J.},
{\bf 313}, 505.

\ref
Bahcall, N.A., \& Soneira, R.M. 1983 \apj 270 20/

\ref
Bahcall, N.A., Batuski, D.J., \& Olowin, R.P. 1988, \apjl 333 13/

\ref
Batuski, D.J., Bahcall, N.A., Olowin, R.P., Burns, J.O. 1989, \apj 341 599/

\ref
Barcons, X. 1991. In {\it The Infrared and Sub-mm Sky after COBE}.
Eds. Signore, M. \& Dupraz, C., in press

\ref
Barcons, X., \& Fabian, A.C. 1988 \mnras 230 189/

\ref
Barcons, X., \& Fabian, A.C. 1989 \mnras 237 119/

\ref
Barcons, X., \& Fabian, A.C. 1990 \mnras 243 366/

\ref
Carrera, F.J., \& Barcons, X. 1992, MNRAS, in press

\ref
Condon, J.J. 1974, \apj 188 279/

\ref
Danese, L., De Zotti, G., Fasano, G., \& Franceschini, A. 1986.
\aa 161 1/

\ref
De Zotti, G., Persic, M., Franceschini, A., Danese, L., Palumbo, G.G.C.,
Boldt, \& E.A., Marshall, F.E. 1990, \apj 351 22/

\ref
Efstathiou, G. \& Rees, M.J. 1988 \mnras 230 5P/

\ref
Fabian, A.C., 1975, \mnras 172 149/

\ref
Giacconi, R. et al.\apj 230 540/

\ref
Gioia, I.M., Maccacaro, T., Schild, R.E., \& Wolter, A. 1990,
\apjs 72 567/

\ref
Griffiths, R.E. et al. 1983 \apj 269 375/

\ref
Gush, H.P., Halpern, M., \& Wishnow, E.H. 1990, \physrevl 65 537/

\ref
Hamilton, T.T., \& Helfand, D.J. 1987, \apj 318 93/

\ref
Jahoda, K., \& Mushotzky, R.F. 1989, \apj 346 638/

\ref
Kaiser, N. 1986\mnras 222 323/

\ref
Kruszewski, A. 1989, in {\it Large Scale Structure and Motions in the
Universe},
ed. M. Mezzetti, G. Giuricin, F. Mardirossian, and M. Ramella (Dordrecht:
Kluwer), p. 385.

\ref
Mc Cammon, D. \& Sanders, W. T. 1990, \araa 28 657/

\ref
Martin-Mirones, J.M., De Zotti, G., Boldt, E.A., Marshall, F.E., Danese,
L., Franceschini, A., \& Persic, M. 1991 \apj 379 507/

\ref
Miyaji, T., \& Boldt, E. 1990, \apjl 327 25/

\ref
Mushotzky, R.F. 1988, in {\it Hot Thin plasma in Astrophysics}, ed R.
Pallavicini (Dordrecht:Reidel), P. 273

\ref
Olivier, S., Blumenthal, G.R., Deel, A., Primack, J.R., \&
Stanhill, D. 1990, \apj 356 1/

\ref
Peebles, P.J.E. 1980, {\it The Large-Scale Structure of the Universe}
(Princeton: Princeton University Press)

\ref
Persic, M., De Zotti, G., Boldt, E.A., Marshall, F.E., Danese, L.,
Franceschini, A., \& Palumbo, G.G.C. 1989 \apjl 336 47/

\ref
Piccinotti, G., Mushotzky, R.F., Boldt, E.A., Holt, S.S., Marshall, F.E.,
Serletmitsos, P.J., \& Shafer, R.A. 1982, \apj 253 485/

\ref
Pye, J.P.,\& Warwick, R.S. 1979, \mnras 187 905/

\ref
Scheuer, P.A.G. 1957, Proc. Cambridge Phil. Soc., 53, 764.

\ref
Scheuer, P.A.G. 1974, \mnras 166 329/

\ref
Schwartz, D.A., Murray, S.S.,\& Gursky, H. 1976 \apj 204 315/

\ref
Sebok, W.L. 1986, \apjs 62 301/

\ref
Shafer, R.A. 1983, Ph. D. thesis, University of Maryland

\ref
Shafer, R.A, \& Fabian, A.C. 1983, in IAU Symposium 104, Early
evolution of the Universe and Its Present Structure, ed. G.O. Abell and
G. Chincarini (Dordrecht:Reidel), p. 333

\ref
Shanks, T., Boyle, B.J., and Peterson, B.A. 1988, in {\it Proceedings
of a Workshop on Optical Surveys for Quasars}, ed. P.S. Osmer, A.C. Porter,
R.F. Green, and C.B. Foltz (San Francisco:Astronomical Society of the
Pacific), p. 244.

\ref
Shaver, P.A. 1988, in {\it IAU Symposium 130, Large Scale Structures of
the Universe}, ed. J. Audouze, M.-C. Pelletan, and A. Szalay (Dordrecht:
Kluwer), p. 359.

\ref
Shaver, P.A., Iovino, A, \& Pierre, M. 1989, in {\it Large Scale Structure
and
Motions in The Universe}, ed M. Mezzetti et al. (Dordrecht:Kluwer), p. 101

\ref
Worrall, D.M., Marshall, F.E., \& Boldt E.A. 1979, \nat 281 127/

\ref
Worrall, D.M., Marshall, F.E. 1984 \apj 276 434/